\documentclass[aps,prd,floatfix,nofootinbib,twocolumn]{revtex4}
\usepackage{graphicx}
\usepackage{amsmath}
\usepackage{amssymb}
\usepackage{subfigure}
\usepackage{bm}
\newcommand{\beq}{\begin{equation}}
\newcommand{\eeq}{\end{equation}}
\newcommand{\bea}{\begin{eqnarray}}
\newcommand{\eea}{\end{eqnarray}}
\newcommand{\MeV}{\;\text{MeV}}

\begin{document}
\title{ Dual meson condensates in the Polyakov-loop enhanced linear sigma model }

\author{Zhao~Zhang}\email{zhaozhang@pku.org.cn}
\author{Haipeng Lu}
\affiliation{ School of Mathematics and Physics, North China
Electric Power University, Beijing 102206, China}

\begin{abstract}

Whether dual meson condensates can indicate deconfinement are investigated in a Polyakov-loop enhanced linear sigma model 
by imposing the twisted boundary conditions. It is confirmed that the rapid rise of the dual sigma condensate with $T$ at 
zero density is driven by the chiral transition, no matter the Polyakov-loop dynamics and or the Dirac-sea contribution 
are included or not. For finite isospin chemical potential $\mu_I>m_\pi/2$, the dual sigma condensate shows abnormal thermal 
behavior which decreases with $T$ below the melting temperature $T_c^{I_3}$ of pion superfluidity; On the other hand, even 
the dual pion condensate always increases with $T$, its maximum slope locates exactly at $T_c^{I_3}$ rather than the 
deconfinement temperature $T_c^{P}$ determined by the Polyakov-loop.
The dual vector meson condensate for $\mu_I>m_\pi/2$ is also calculated. This quantity is more sensitive to the chiral
transition when taking into account the Dirac-sea contribution. Our study further suggests that it should be cautious
to use the dual observables as indicators of deconfinement, at least in QCD models.

\vspace{10pt} PACS number(s): 12.38.Aw; 11.30.RD; 12.38.Lg;
\end{abstract}

\date{\today}
\maketitle

\section{Introduction}

Understanding the confinement-deconfinement phase transition at finite temperature and density is a very
important task in high energy nuclear physics. However, it is conceptually difficult to define a relevant
order parameter in QCD. So far, how to describe the deconfinement transition is still a subtle problem.

In the heavy quark limit, the expectation value of the Polyakov-loop (PL) is the true order parameter for
deconfinement, which is directly related to the center symmetry. Usually, PL is also extensively used to
indicate the quark deconfining transition in lattice QCD (LQCD) \cite{Aoki:2009sc,Borsanyi:2010bp,Bazavov:2013yv,Bazavov:2016uvm}
and effective models \cite{Fukushima:2017csk}, even though the center
symmetry is badly broken by light quarks.
Besides PL, some other quantities or criteria are also proposed and used to determine the
deconfinement transition in the literature. These include the QCD-monopole in the dual Ginzburg-Landau
theory \cite{Suganuma:1995mn}, the center vortex \cite{Chernodub:2011pr}, the PL fluctuation \cite{Lo:2013hla},
the entropy in the framework of a hybrid model \cite{Miyahara:2016din}, the quark number holonomy based on the
topological picture \cite{Kashiwa:2016vrl,Kashiwa:2017yvy} and so on.

Recently, the dressed PL (DPL) is suggested as an appropriate order parameter for deconfinement
in QCD \cite{Bilgici:2008qy,Bilgici:2009tx}. This quantity is defined as the first Fourier moment of
the quark condensate obtained under the twisted boundary condition for fermions. In lattice language,
DPL includes contributions of infinite closed loops with winding number one around the temporal direction.
Thus it transforms in the same manner as PL under the center transformation
(PL only includes the shortest loop contribution). For infinite quark masses, DPL reduces
to PL since the spacial fluctuations are suppressed \footnote{One can construct many dual
observables which belong to the same class as PL under the center transformation
\cite{Braun:2009gm,Zhang:2010ui}.}.

One merit of DPL is that it interpolates between the quark condensate and PL, which may imply some intrinsic
relation between chiral transition and deconfinement. Another is that it can also be calculated in some QCD
models. The previous investigations in LQCD \cite{Bilgici:2008qy,Bilgici:2009tx,Zhang:2010ui}, the truncated
Dyson-Schwinger equations (tDSE) \cite{Fischer:2009gk,Fischer:2009wc,Fischer:2010fx,Fischer:2011mz} and
Nambu-Jona-Lasinio (NJL) type models \cite{Kashiwa:2009ki,Gatto:2010qs,Mukherjee:2010cp} all indicate DPL
exhibits the order parameter-like behavior, just as PL. The coincidence of the two phase transitions, namely
$T_c^\chi{\approx}T_c^D$, is obtained in these studies.

Since the center symmetry is seriously broken, one may ask to what extent DPL can indicate deconfinement in QCD.
Model studies may shed some light on this question. A particularly noticeable calculation \cite{Mukherjee:2010cp}
is that DPL obtained in NJL is very similar to the lattice result: it increases with $T$ and changes rapidly near
$T_c^\chi$. Since NJL has no gluon fields, such a rapid rise should have little relation with center symmetry.
Actually, a Ginzburg-Landau analysis \cite{Benic:2013zaa} manifests that it is totally driven by chiral transition.
Subsequent studies \cite{Marquez:2015bca,Zhang:2015baa} using NJL variants with different confining elements
obtain the similar conclusion. It is found that the rapid rise of DPL has no effect on the change of confining
properties of the quark propagator \cite{Marquez:2015bca}. By considering gluon degrees of freedom with center
symmetry, it is confirmed that the rapid rise of DPL is still determined by chiral restoration rather than the
increase of PL in \cite{Zhang:2015baa}, where PNJL \cite{Fukushima:2003fw,Roessner:2006xn} is used
\footnote{Dual quark condensates in PNJL are first calculated by Kashiwa in \cite{Kashiwa:2009ki}, where
the role of vector interaction is addressed.}. Moreover, Ref.~\cite{Zhang:2015baa} shows that for $\mu_I>m_\pi/2$,
the dual pion condensate (DPC) behaves similarly as PL, while DPL decreases with $T$ until the pion
condensate melts away. All these suggest that DPL calculated in NJL type models should not be regarded as
the deconfinement order parameter.

This raises a question: Whether DPL is merely sensitive to the chiral transition in QCD? If so, using this
quantity to conclude the coincidence of chiral restoration and deconfinement should be problematic. In this sense,
it is necessary to first check whether the above NJL conclusion also holds in other QCD models, especial those
with hadron degrees of freedom.  The main purpose of this work is to try to extract DPL in the PL augmented
linear sigma model (PL$\sigma$M) of QCD (also known as PQM) and compare it with the NJL results
in \cite{Benic:2013zaa,Marquez:2015bca,Zhang:2015baa}.

PL$\sigma$M \cite{Schaefer:2007pw} is a popular chiral model which has been extensively used to explore the QCD
phase transitions. Different from PNJL, this model includes three types of degrees of freedom: quarks, mesons, and
gluons. The philosophy of PL$\sigma$M is that quarks and gluons are relevant objects for $T>T_c$, while mesons
play the dominant role in low temperatures. Compared to NJL, the L$\sigma$M part in PL$\sigma$M has the merit
of renormalizability. It is argued that PL$\sigma$M is more suitable to study the QCD phase diagram than PNJL at
low baryon density \cite{Fukushima:2017csk}. In the literature, (P)L$\sigma$M is also frequently employed to study
the inhomogeneous chiral condensates at high baryon density \cite{Buballa:2014tba} and the chiral transition in a
magnetic field \cite{Andersen:2014xxa}. However, this model is seldom used to study the physics at imaginal
chemical potential.

Since L$\sigma$M can be viewed as a partially bosonized version of NJL in a certain sense, the dual observables
related to some quark bilinears may be assessed indirectly through studying the corresponding meson condensates in
(P)L$\sigma$M by imposing the twisted boundary conditions. In this article, DPL and DPC mentioned above are
evaluated in PL$\sigma$M by researching the dual sigma and pion condensates at the mean field level. Beyond
\cite{Zhang:2015baa}, the dual vector meson condensate related to the isospin density is also calculated.

Unlike (P)NJL, the Dirac-sea contribution is not necessary for the dynamical chiral symmetry breaking in (P)L$\sigma$M. There
exists subtlety on how to treat this term in (P)L$\sigma$M, which is ignored in \cite{Scavenius:2000qd,Schaefer:2007pw} but
taken into account in \cite{Skokov:2010sf}. In our calculation, both treatments are adopted and compared. The paper is organized
as follows. In Sec.II, the dual meson condensates for both zero $\mu_I$ and $\mu_I>m_\pi/2$ in PL$\sigma$M are introduced, where
the twisted boundary condition is used. The numerical results and discussion are given in Sec.III. In Sec.IV, we summarize.

\section{ Dual condensates in PL$\sigma$M with twisted boundary condition }

\subsection{ Two flavor PL$\sigma$M at finite $\mu$ and $\mu_I$ }

We adopt the following lagrangian density of the two-flavor PL$\sigma$M \cite{Ueda:2013sia}

\begin{eqnarray}
{\mathcal{L}}&=&\bar{q} S_0^{-1}q+\frac{1}{2}{\left({{\partial_\mu }\sigma}\right)^2}+\frac{1}{2}{\left({{\partial_\mu}\vec\pi} \right)^2}
-U\left( {\sigma ,\vec\pi} \right)\nonumber\\
&&-\frac{1}{4}\omega^{\mu\nu}\omega_{\mu\nu}-\frac{1}{4}\vec{R}^{\mu\nu}\vec{R}_{\mu\nu}+\frac{1}{2}m^2_{v}(\omega^\mu\omega_\mu+\vec{R}^\mu\cdot\vec{R}_\mu)\nonumber\\
&&-\mathcal{U}\left(\Phi,\bar{\Phi},T\right)\label{PQM}
\end{eqnarray}
with
\begin{equation}
S_0^{-1}={i\gamma^\mu}{D_\mu}-g\left({\sigma+i{\gamma_5}\vec{\tau}\cdot\vec\pi} \right)
-g_{\omega}{\gamma^{\mu}}\omega_{\mu}-g_{\rho}{\gamma^{\mu}}\vec{\tau}\cdot{\vec{R}_\mu},
\end{equation}
where $q$ denotes the quark field, $\vec\tau$ is the Pauli matrix in the flavor space. The mesonic potential $U$ is
given as
\begin{equation}
U(\sigma,\vec{\pi})=\lambda(\sigma^2+\vec\pi^2-v^2)^2/4-h\sigma,
\end{equation}
therein $\sigma$ and $\vec\pi$ are the isoscalar-scalar and isovector-pseudoscalar meson fields. The vector meson degrees of
freedom are also taken into account and $\omega^{\mu\nu}$ and $\vec{R}^{\mu\nu}$ are the field tensors of $\omega$ and $\rho$
mesons, respectively. The term $\mathcal{U}\left(\Phi,\bar{\Phi},T\right)$ is the PL potential, which respecting the $Z(3)$
center symmetry.

In PL$\sigma$M, the quark chemical potential $\mu$ is introduced in the same way as in QCD. However, the introduction
of $\mu_I$ is quite different. Under the isospin $U(1)_{I_3}$ transformation, the quark and pion fields change
in the following way:
\begin{equation}
q\longrightarrow e^{i\tau_3\theta}q,\quad q^{\dag}\longrightarrow e^{-i\tau_3\theta}q^{\dag},\quad
\pi_{\pm}\longrightarrow e^{{\mp}i2\theta}\pi_{\pm}.
\end{equation}
The corresponding conserved current takes the form
\begin{equation}
J^{3}_\mu=\bar{q}\tau_3\gamma_\mu{q}+2i(\pi_-\partial_\mu\pi_+-\pi_+\partial_\mu\pi_-),
\end{equation}
where
\begin{equation}
\pi_\pm=\pi_1\pm{i\pi_2}.
\end{equation}
Accordingly, $\mu_I$ can be introduced by adding the term $\mu_II^3$
to the Hamiltonian, where the associated conserved charge is
\begin{equation}
 I^{3}=\int{d^3\vec{x}}(\bar{q}\tau_3\gamma_0{q}+\pi_1\partial_t\pi_2-\pi_2\partial_t\pi_1).
\end{equation}

The lagrangian density \eqref{PQM} is then modified at finite $\mu$ and $\mu_I$ by the following
replacements
\begin{equation}
S_0^{-1}\rightarrow S_0^{-1}+\gamma_0\widehat{\mu},\label{change1}
\end{equation}
and
\begin{equation}
{\left({{\partial_\mu}\vec\pi} \right)^2}\rightarrow
{\left({{\partial_\mu}\pi_0} \right)^2}+(({{\partial_\mu}+2\mu_I\delta^0_\mu)\pi_+})({{\partial_\mu}+2\mu_I\delta^0_\mu)\pi_-},\label{change2}
\end{equation}
where
\begin{equation}
\hat{\mu}=\bigg(\begin{array}{cc}
    \mu_u & \\
     & \mu_d\end{array}
 \bigg)=\bigg(\begin{array}{cc}
    \mu+\mu_I & \\
     & \mu-\mu_I\end{array}
 \bigg).
\end{equation}
The reason for the appearance of $\mu_I^2\pi_+\pi_-$ in \eqref{change2} is that the generalized momentums of
pion fields has been integrated out according to the standard derivation \cite{Kapusta:book}.

The phase diagram of a two-flavor L$\sigma$M at finite $\mu$, $\mu_I$ and $T$ has been investigated in \cite{Kamikado:2012bt},
where the pion superfluid phase is also studied. Note that the effects of the PL dynamics and vector mesons are all ignored
in that work. Taking into account these elements and following the treatment in \cite{Zhang:2006gu}, we
derive the mean field thermal potential of PL$\sigma$M at finite $\mu$
and $\mu_I$
\begin{eqnarray}\label{thermalp}
&&\Omega=-2N_c\int{\frac{d^3p}{(2\pi)^3}}\big[E_p^-+E_p^+\big]\theta(\Lambda^2-\vec{p}^2)\nonumber\\
&&-2T\int{\frac{d^3p}{(2\pi)^3}}\Big\{\ln\Big[1+3\big(\Phi+\bar{\Phi}\mathrm{e}^{-\left(E_p^--\mu'\right)\beta}\big)\mathrm{e}^{-\left(E_p^--\mu'\right)\beta}
\nonumber\\
&&+\mathrm{e}^{-3\left(E_p^--\mu'\right)\beta}\Big]+\ln\Big[1+3\big(\bar{\Phi}+\Phi\mathrm{e}^{-\left(E_p^-+\mu'\right)\beta}\big)\mathrm{e}^{-\left(E_p^-+\mu'\right)\beta}
\nonumber\\
&&+
\mathrm{e}^{-3\left(E_p^-+\mu'\right))\beta}\Big]+\ln\Big[1+3\big(\Phi+\bar{\Phi}\mathrm{e}^{-\left(E_p^+-\mu'\right)\beta}\big)\mathrm{e}^{-\left(E_p^+-\mu'\right)\beta}
\nonumber\\
&&+
\mathrm{e}^{-3\left(E_p^+-\mu'\right)\beta}\Big]+\ln\Big[1+3\big(\bar{\Phi}+\Phi\mathrm{e}^{-\left(E_p^++\mu'\right)\beta}\big)\mathrm{e}^{-\left(E_p^++\mu'\right)\beta}
\nonumber\\
&&+
\mathrm{e}^{-3\left(E_p^++\mu'\right)\beta}\Big]\Big\}-2\mu_I\pi^2-\frac{1}{2}(M_\omega^2\omega^2+M_\rho^2R^2)
\nonumber\\
&&
+U\left( {\sigma, \pi}\right)+{\cal{U}}(\Phi,\bar{\Phi},T),\label{thermp}
\end{eqnarray}
with the quasi particle energy $E_p^{\pm}=\sqrt{(E_p\pm\mu_I')^2+N^2}$ and $E_p=\sqrt{\vec{p}^2+M^2}$
in which the two energy gaps are defined as
\begin{eqnarray}
M=g\sigma,\\
N=g\pi.
\end{eqnarray}
Here (also in the following) $\sigma$ and $\pi$ refer to the vacuum expectation values (VEVs) of the sigma and charged pion mesons
and the later is defined as
\begin{eqnarray}
\pi=\langle{\pi_{+}\rangle}e^{i\theta'}=\langle{\pi_{-}\rangle}e^{-i\theta'}.
\end{eqnarray}
Nonzero $\pi$ indicates the spontaneous breaking of the $U(1)_{I_3}$ symmetry and the phase factor $\theta'$ is the breaking
direction. $\mu'$ and $\mu_I'$ are the shifted quark and isospin chemical potentials
\begin{eqnarray}
\mu'=\mu-g_\omega\omega,\quad
\mu_I'=\mu_I-g_\rho{\rho},\label{shiftedcp}
\end{eqnarray}
where $\omega$ and $\rho$ denote the VEVs of $\omega$ and $\rho_0$ mesons
\begin{eqnarray}
\omega=\langle{\omega_0}\rangle,\quad \rho=\langle{R_0^3}\rangle,
\end{eqnarray}
respectively.

In this paper, we only consider the situation with finite $\mu_I$ and vanishing $\mu$. In this case, $\Phi$ equals to $\bar{\Phi}$
strictly \cite{Zhang:2006gu} and it is free from the sign problem even in the lattice simulation. The reason for the later is that
$\tau_2{\gamma_5}{D}{\gamma_5}\tau_2={D}^\dagger$ which ensures $\text{det}D\geq{0}$\cite{Son:2001}, where $D$ is the Dirac operator.
Minimizing the
thermal dynamical potential (\ref{thermalp}),
the motion equations for the mean fields $\sigma$, $\pi$, $\Phi$ and $\rho$ are determined by
\begin{equation}
\frac{\partial\Omega}{\partial\sigma}=0,\quad\frac{\partial\Omega}{\partial\pi}=0,
\quad\frac{\partial\Omega}{\partial\Phi}=0,\quad\frac{\partial\Omega}{\partial{\rho}}=0.\label{coupleeqs1}
\end{equation}
This set of equations is then solved for the fields $\sigma$, $\pi$, $\Phi$ and $\rho$ as functions of $T$ and $\mu_I$.

\subsection{ Two flavor PL$\sigma$M at finite $\mu_I$ with twisted boundary condition }

To calculate the dual observables, we must adopt the twisted boundary condition in time direction for quarks
\beq
q(x,\beta=1/T)=e^{i\phi}q(x,0)\label{twistedbc},
\eeq
where $\phi$ ranges from zero to $2\pi$. Under this condition, the modified quark chemical potential
$\mu'$ in \eqref{thermalp} should be replaced by $iT(\phi-\pi)$ \cite{Bilgici:2008qy,Braun:2009gm,Kashiwa:2009ki},
which is nothing but an imaginary baryon chemical potential.  There is no sign problem
for purely imaginary baryon chemical potential since
\begin{equation}
\text{det}{D(\mu)}=\text{det}[\gamma_5{D(\mu)}\gamma_5]=\text{det}^*{D(-\mu^*)}.
\end{equation}
For details on lattice simulations at finite $\mu_I$ and imaginal $\mu$, please refer to \cite{Cea:2012ev,DElia:2009bzj}.

Strictly speaking, $\mu'$ at $\phi\neq\pi$ should contain an imaginal part $g_\omega\omega$, even $\mu$ is zero
\footnote{In PL$\sigma$M, $\omega$ is closely related to the dual density proposed in \cite{Braun:2009gm}.}.
Such a term is ignored in our calculation. It has been shown in \cite{Kashiwa:2009ki} that a similar term in PNJL has
little effect on DPL near $T_c^\chi$.  Note that $\mu_I'$ is always real because the imaginary parts of $\mu_u'$ and
$\mu_d'$ cancel each other out. This means $\rho$ is still real for $\phi\neq\pi$. This quantity resembles the isospin
density in NJL with vector interactions \cite{Zhang:2015baa}.

In the standard definition of DPL \cite{Bilgici:2008qy,Bilgici:2009tx}, the twisted boundary condition is imposed
on the Dirac operator $D_{\phi}$, and the bracket $\langle\cdot\cdot\cdot\rangle$ still keeps the antiperiodic
condition with $\phi=\pi$. So in our calculation, $\Phi$ as a function of $T$ and $\mu_I$ is first obtained by
solving \eqref{coupleeqs1} using the physical boundary condition. The other quantities, such
as $\sigma(\phi)$, $\pi(\phi)$ and $\rho(\phi)$ are then determined by the following coupled equations
\begin{equation}
\frac{\partial\Omega}{\partial\sigma(\phi)}=0,\quad\frac{\partial\Omega}{\partial\pi(\phi)}=0,
\quad\frac{\partial\Omega}{\partial{\rho}(\phi)}=0,\label{coupleeqs2}
\end{equation}
with $\Phi$ keeping its value for $\phi=\pi$. Such a treatment is consistent with
\cite{Kashiwa:2009ki,Zhang:2015baa}.

\subsection{ Dual meson condensates at finite $\mu_I$}

According to \cite{Bilgici:2008qy}, DPL is defined as
\beq
\Sigma^{(1)}_{\sigma}=-\int^{2\pi}_0\frac{d\phi}{2\pi}e^{-i\phi}\langle{\bar{q}q\rangle}_{\phi}\label{dualquarkc},
\eeq
where $\langle{\bar{q}q\rangle}_{\phi}$ is the generalized quark condensate.
Similarly, the dual pion condensate
\beq
\Sigma^{(1)}_{\pi}=-\int^{2\pi}_0\frac{d\phi}{2\pi}e^{-i\phi}\langle{\bar{q}i\gamma_5\tau_1{q}}\rangle_{\phi}\label{dualpionc}
\eeq
is introduced in \cite{Zhang:2015baa}.
Both (and also the dual density proposed in \cite{Braun:2009gm}) are gauge invariant, which merely including contributions
of closed loops with wingding number one. As mentioned, they belong to the same class as PL under the $Z(3)$ center
transformation.

Following \eqref{dualquarkc} and \eqref{dualpionc}, we can construct the dual sigma condensate (D$\sigma$C)  and the dual
pion condensate (D$\pi$C) in PL$\sigma$M, namely
\beq
\Sigma^{1}_{\sigma}=\int^{2\pi}_0\frac{d\phi}{2\pi}e^{-i\phi}{\sigma}({\phi})\label{dqcPQM},
\eeq
and
\beq
\Sigma^{1}_{\pi}=\int^{2\pi}_0\frac{d\phi}{2\pi}e^{-i\phi}{\pi}({\phi})\label{dpcPQM}.
\eeq
Since the VEVs of meson fields are gauge invariant, the first moments of $\sigma(\phi)$ and
$\pi(\phi)$ also belong to the same class as PL under the center transformation. Evidently, D$\sigma$C and D$\pi$C
correspond to $\Sigma^{(1)}_{\sigma}$ and $\Sigma^{(1)}_{\pi}$, respectively. The main task of this work is to test
whether these dual meson condensates could be used as order parameters for deconfinement in PL$\sigma$M.

Besides D$\sigma$C and D$\pi$C, we can also define the dual vector meson condensate (D$\rho$C) in PL$\sigma$M, namely
\beq
\Sigma^{1}_{\rho}=-\int^{2\pi}_0\frac{d\phi}{2\pi}e^{-i\phi}{\rho}({\phi})\label{dvcPQM}.
\eeq
This quantity is nonzero at finite $T$ and $\mu_I$ (or zero $T$ for $\mu_I>m_\pi/2$). As mentioned, $\rho(\phi)$
corresponds to the isospin density $\langle\bar{q}\gamma_0\tau_3q\rangle_{\phi}$ in QCD or NJL with vector interactions.
In this sense, $\Sigma^{1}_{\rho}$ is analogous to the dual density proposed in \cite{Braun:2009gm}. It is interesting
to check whether this dual isospin density can be used to indicate the deconfinement transition.

\subsection{ Model parameters }

Two sets of model parameters related to the (pseudo-) scalar mesons are used in our calculations.
The first is adopted from \cite{Scavenius:2000qd}, where the fermion vacuum contribution is ignored.
Namely, $g=m_q/f_\pi$, $\lambda=(m_\sigma^2-m_\pi^2)/(2f_\pi^2)$, $v^2=f_\pi^2-h/(\lambda{f_\pi})$, and
$h=m_\pi^2f_\pi$ with $m_\pi=138\MeV$, $m_\sigma=600\MeV$, $\langle{\sigma}\rangle\equiv{f_\pi}=93\MeV$
and $m_q=300\MeV$. The second is taken from \cite{Skokov:2010sf}, where the Dirac-sea contribution
is included and the momentum cutoff $\Lambda=600\MeV$ is used. The parameters $g$, $\lambda$, $v$ are fixed as
$g=3.60$, $\lambda=3.06$, and $v^2=-(578)^2\MeV^2$ respectively.

The same parameters related to vector mesons are used in both cases. For simplicity, we assume $g_\omega=g_\rho$
\footnote{In general, $g_\omega$ is different from $g_\rho$ which may leads to flavor mixing at finite $\mu_I$\cite{Zhang:2013oia}.},
which is fixed as $0.25g$. We have checked that our main conclusion is insensitive to $g_\rho$. We also assume
the common vector meson masses ($m_\omega=m_\rho=770 \MeV$).
The logarithm PL potential \cite{Roessner:2006xn} is adopted. It has been reported that this type of $\mathcal{U}$
can reproduce the LQCD data at finite imaginary chemical potential, but the polynomial one does
not \cite{Kashiwa:2009ki}. Following \cite{Kashiwa:2009ki,Zhang:2015baa}, the parameter $T_0$ in the logarithm
potential is fitted as $200 \MeV$.

\begin{figure}[t]
\hspace{-.0\textwidth}
\begin{minipage}[t]{.45\textwidth}
\includegraphics*[width=\textwidth]{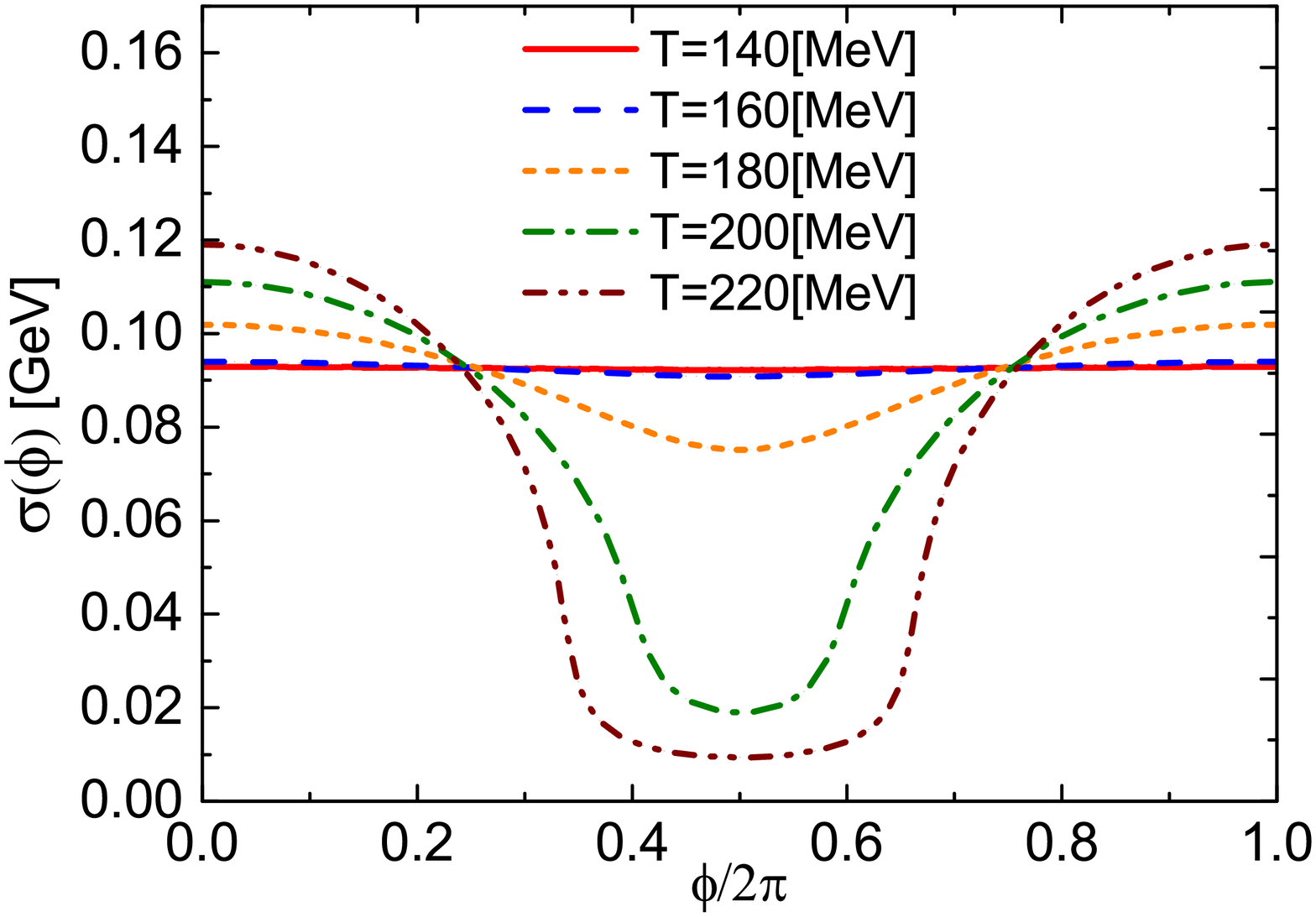}
\centerline{(a)}
\end{minipage}
\hspace{-.05\textwidth}
\begin{minipage}[t]{.45\textwidth}
\includegraphics*[width=\textwidth]{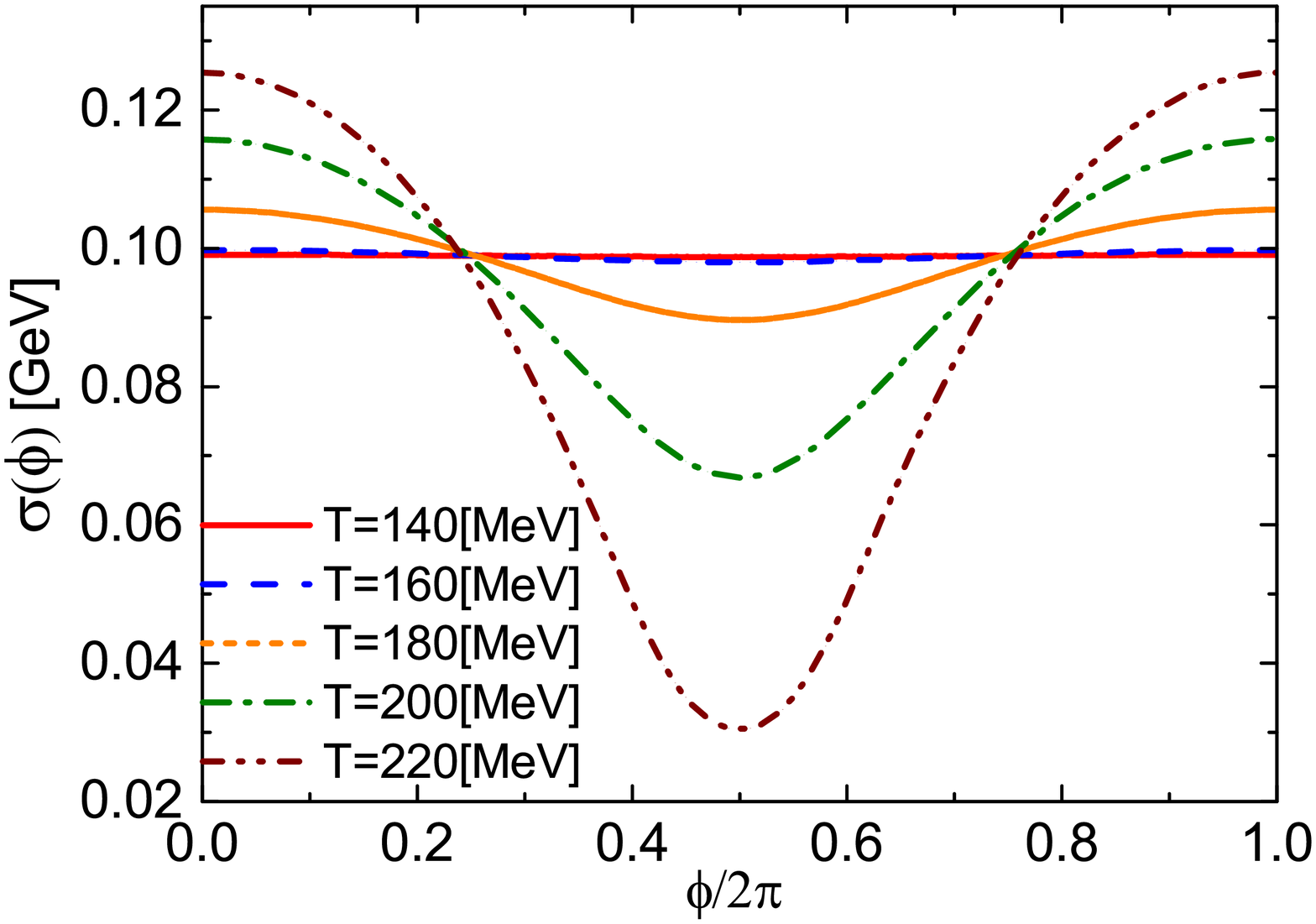}
\centerline{(b) }
\end{minipage}
\caption{ Twisted angle dependence of the generalised sigma condensate at different temperatures for zero $\mu_I$
in PL$\sigma$M. The upper (lower) panel is obtained by ignoring (considering) the Dirac-sea contribution. }
\label{fig:sigmazeromi}
\end{figure}

\begin{figure}[t]
\hspace{-.0\textwidth}
\begin{minipage}[t]{.45\textwidth}
\includegraphics*[width=\textwidth]{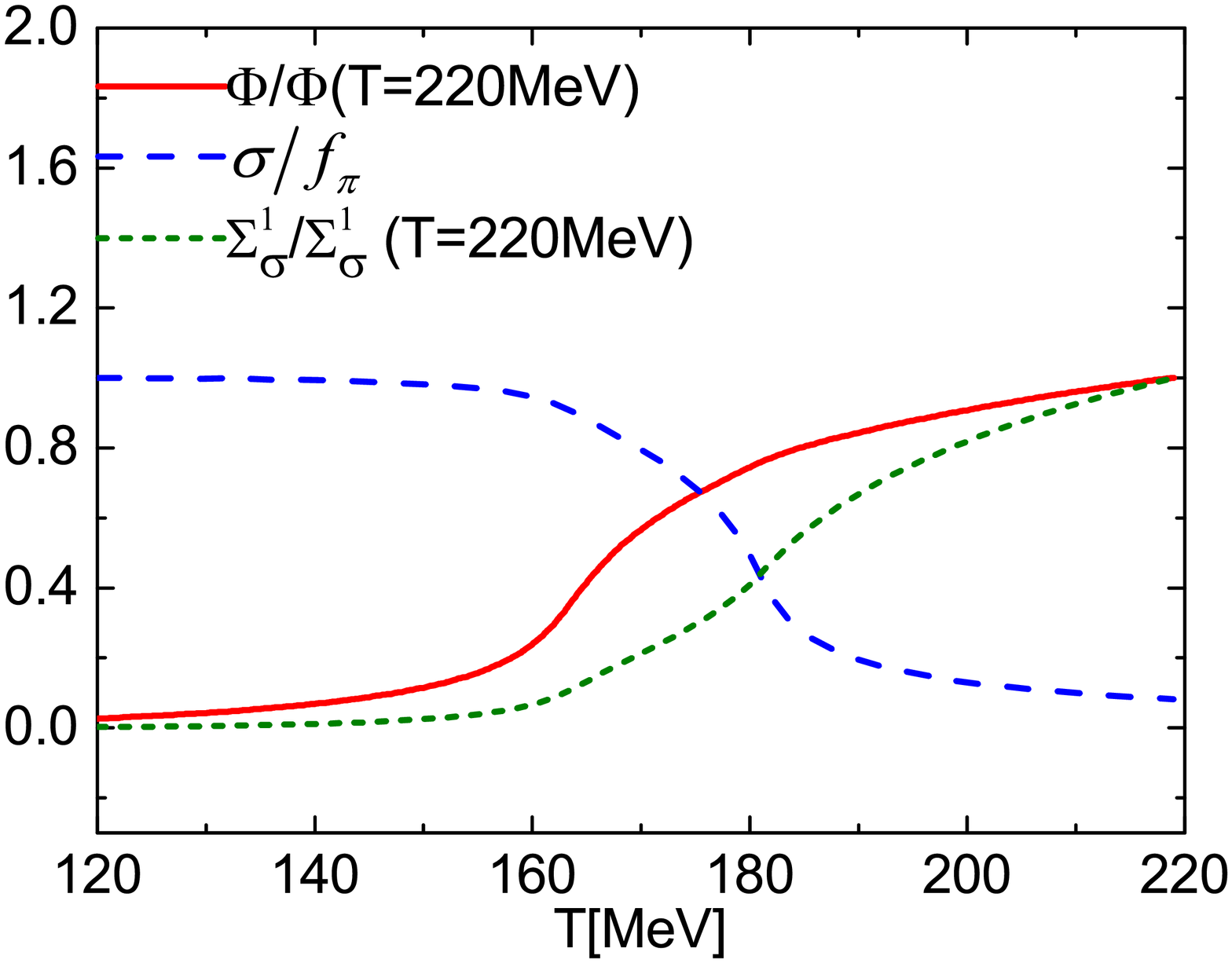}
\centerline{(a) }
\end{minipage}
\hspace{-.05\textwidth}
\begin{minipage}[t]{.45\textwidth}
\includegraphics*[width=\textwidth]{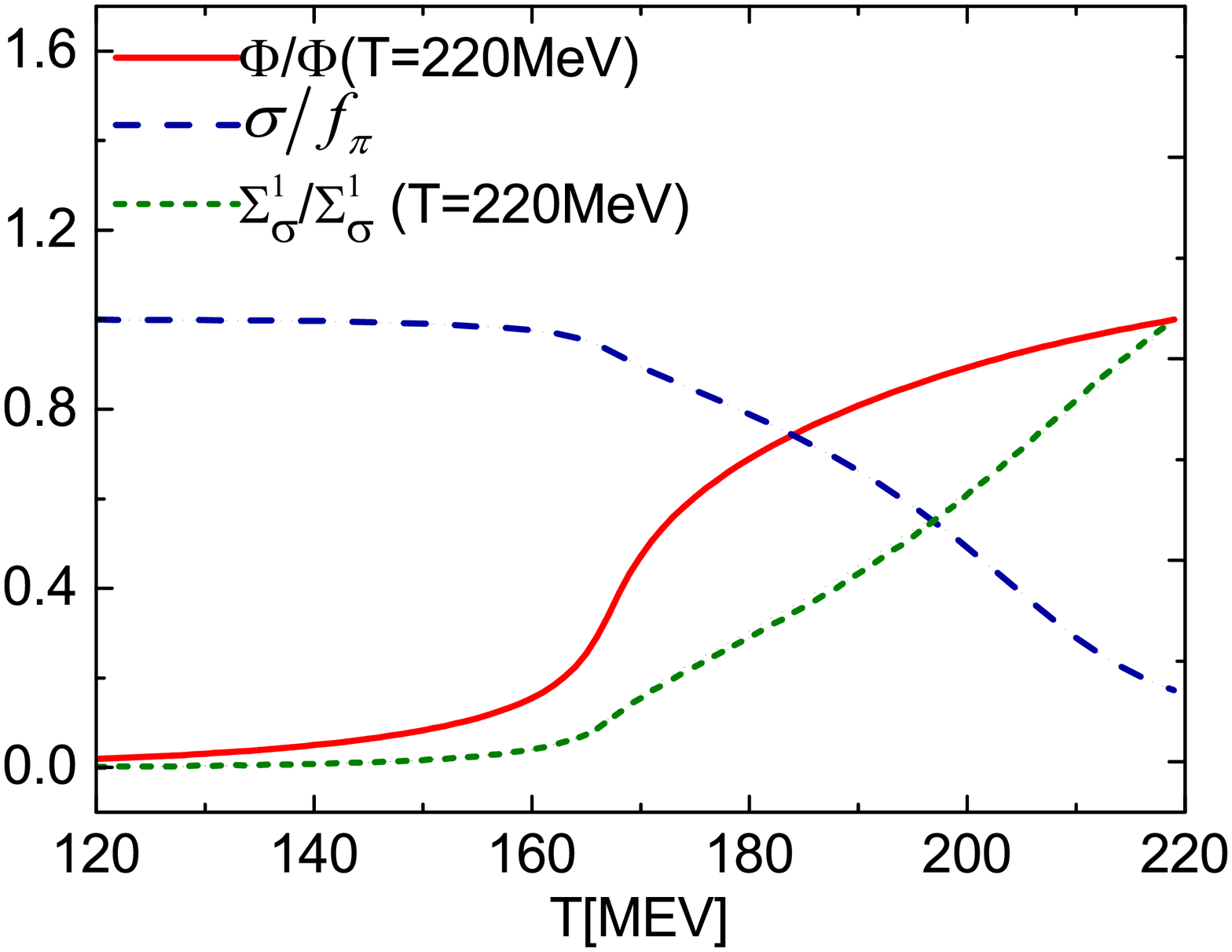}
\centerline{(b) }
\end{minipage}
\caption{ The normalized PL, sigma condensate and  dual sigma condensate as functions of $T$ for
zero $\mu_I$ in PL$\sigma$M. The upper (lower) panel is obtained by ignoring (considering) the Dirac-sea contribution. }
\label{fig:dualzeromi}
\end{figure}

\begin{figure}[t]
\hspace{-.0\textwidth}
\begin{minipage}[t]{.45\textwidth}
\includegraphics*[width=\textwidth]{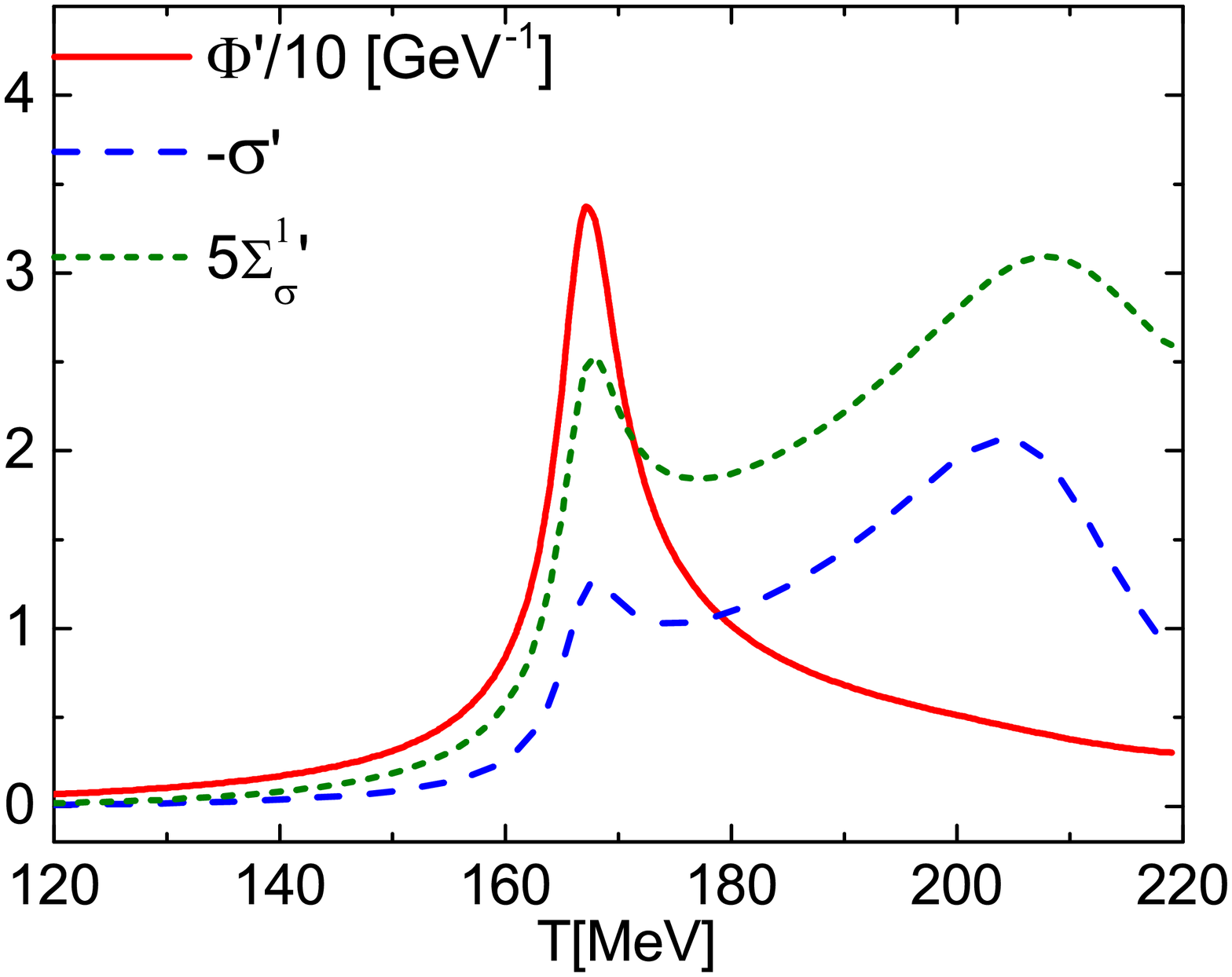}
\centerline{(a) }
\end{minipage}
\hspace{-.05\textwidth}
\begin{minipage}[t]{.45\textwidth}
\includegraphics*[width=\textwidth]{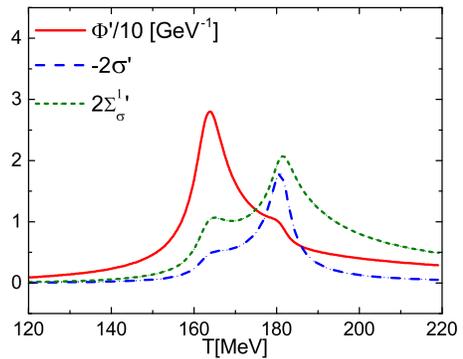}
\centerline{(b) }
\end{minipage}
\caption{ T-derivatives of the PL, sigma condensate and dual sigma condensate as functions of $T$ for
zero $\mu_I$ in PL$\sigma$M. The upper (lower) panel is obtained by ignoring (considering) the Dirac-sea contribution. }
\label{fig:suszeromi}
\end{figure}

\begin{figure}[t]
\hspace{-.0\textwidth}
\begin{minipage}[t]{.45\textwidth}
\includegraphics*[width=\textwidth]{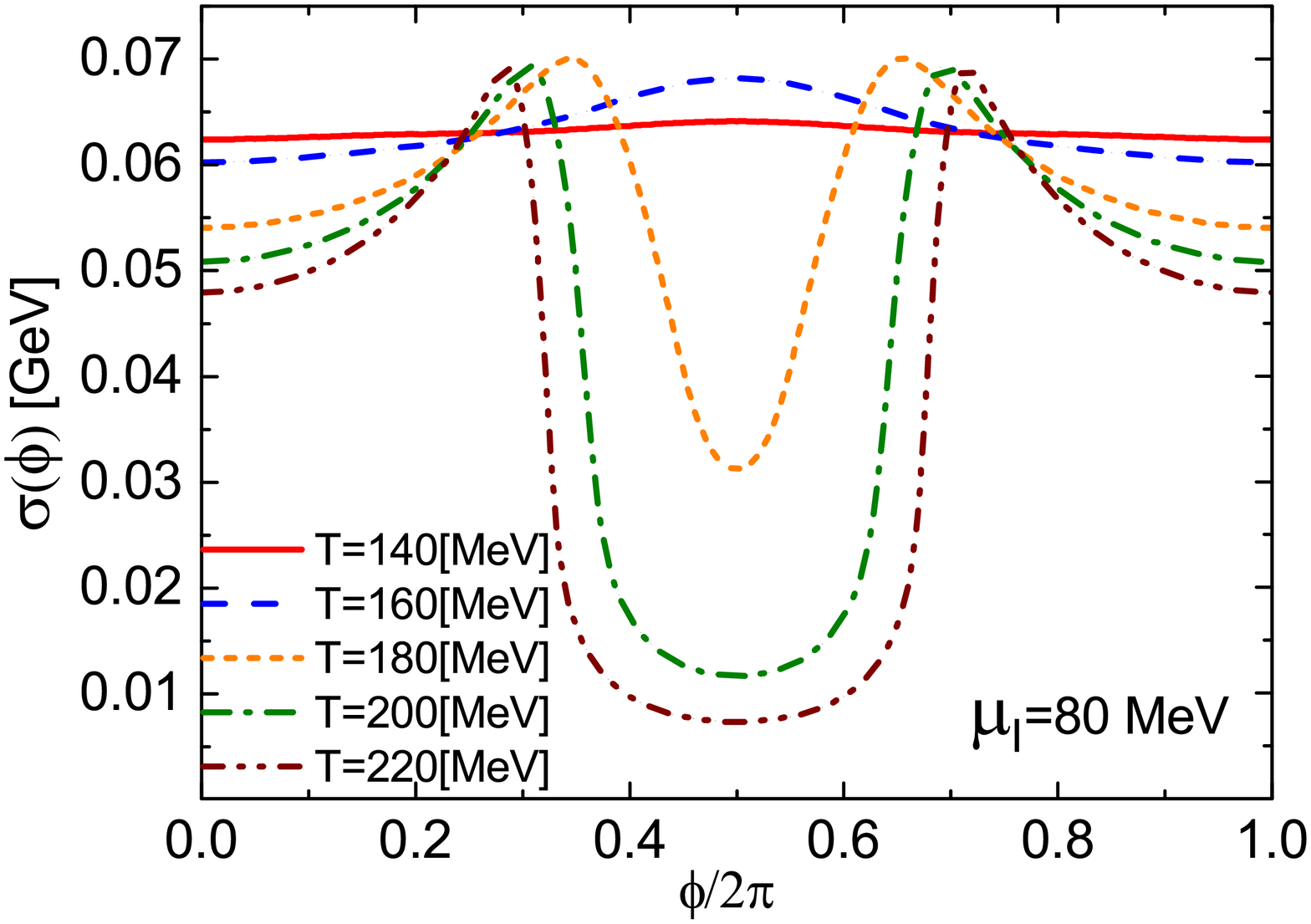}
\centerline{(a) }
\end{minipage}
\hspace{-.05\textwidth}
\begin{minipage}[t]{.45\textwidth}
\includegraphics*[width=\textwidth]{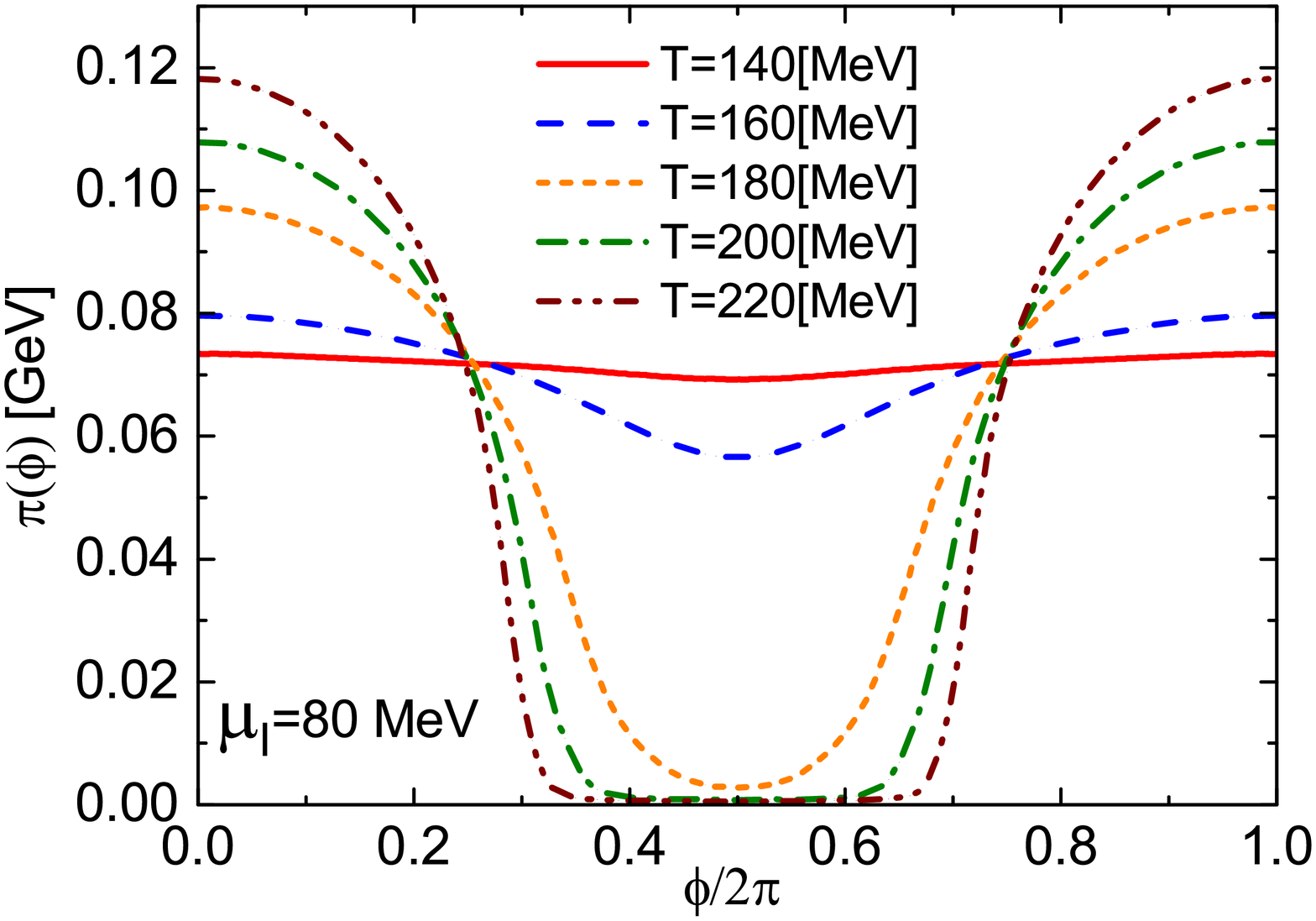}
\centerline{(b) }
\end{minipage}
\hspace{-.05\textwidth}
\begin{minipage}[t]{.45\textwidth}
\includegraphics*[width=\textwidth]{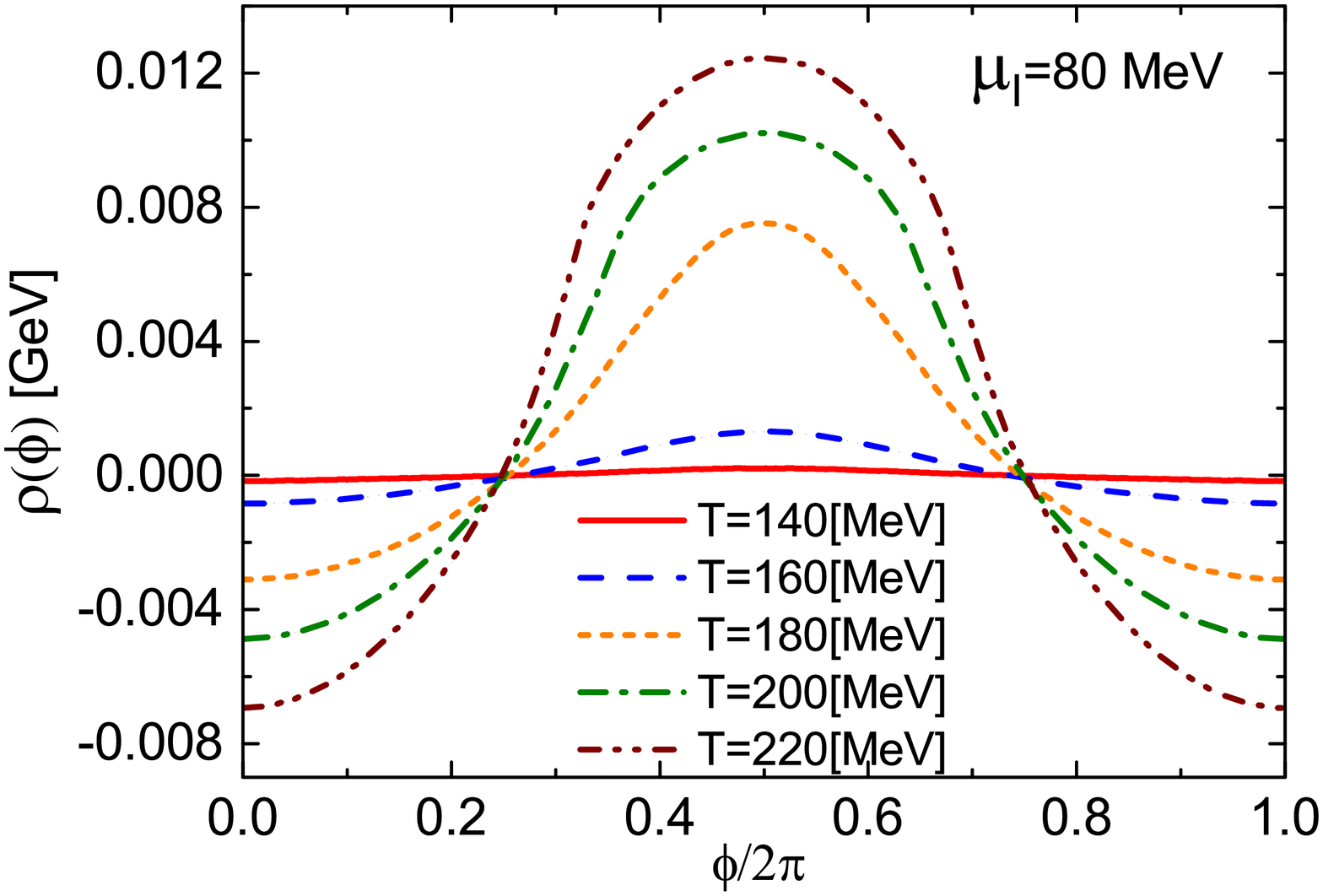}
\centerline{(c) }
\end{minipage}
\caption{ The twisted angle dependences of $\sigma$, $\pi$ and $\rho$ at $\mu_I=80 \MeV$
for different temperatures in PL$\sigma$M. The Dirac-sea contribution is ignored. }
\label{fig:dualc1}
\end{figure}

\begin{figure}[t]
\hspace{-.0\textwidth}
\begin{minipage}[t]{.45\textwidth}
\includegraphics*[width=\textwidth]{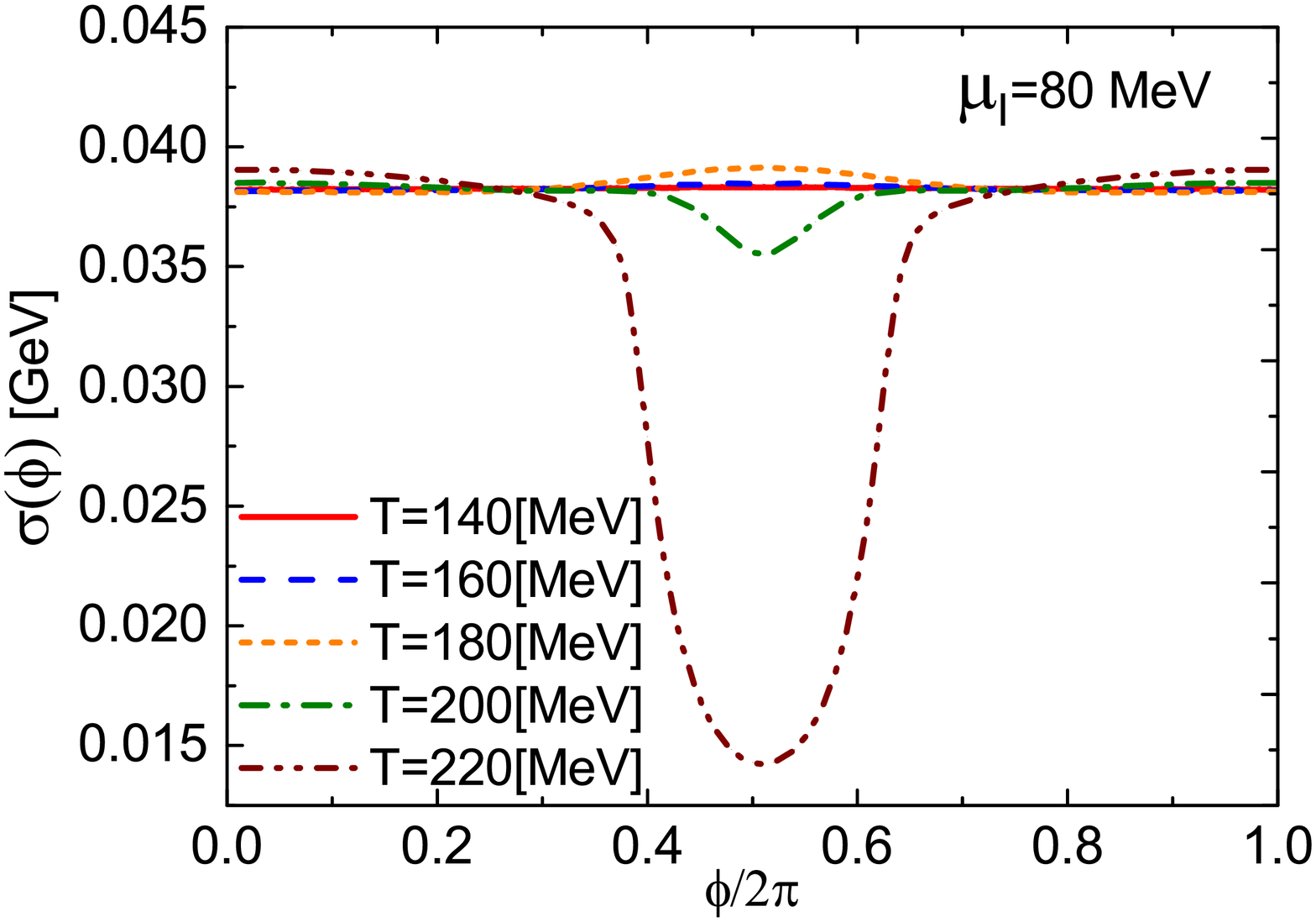}
\centerline{(a) }
\end{minipage}
\hspace{-.05\textwidth}
\begin{minipage}[t]{.45\textwidth}
\includegraphics*[width=\textwidth]{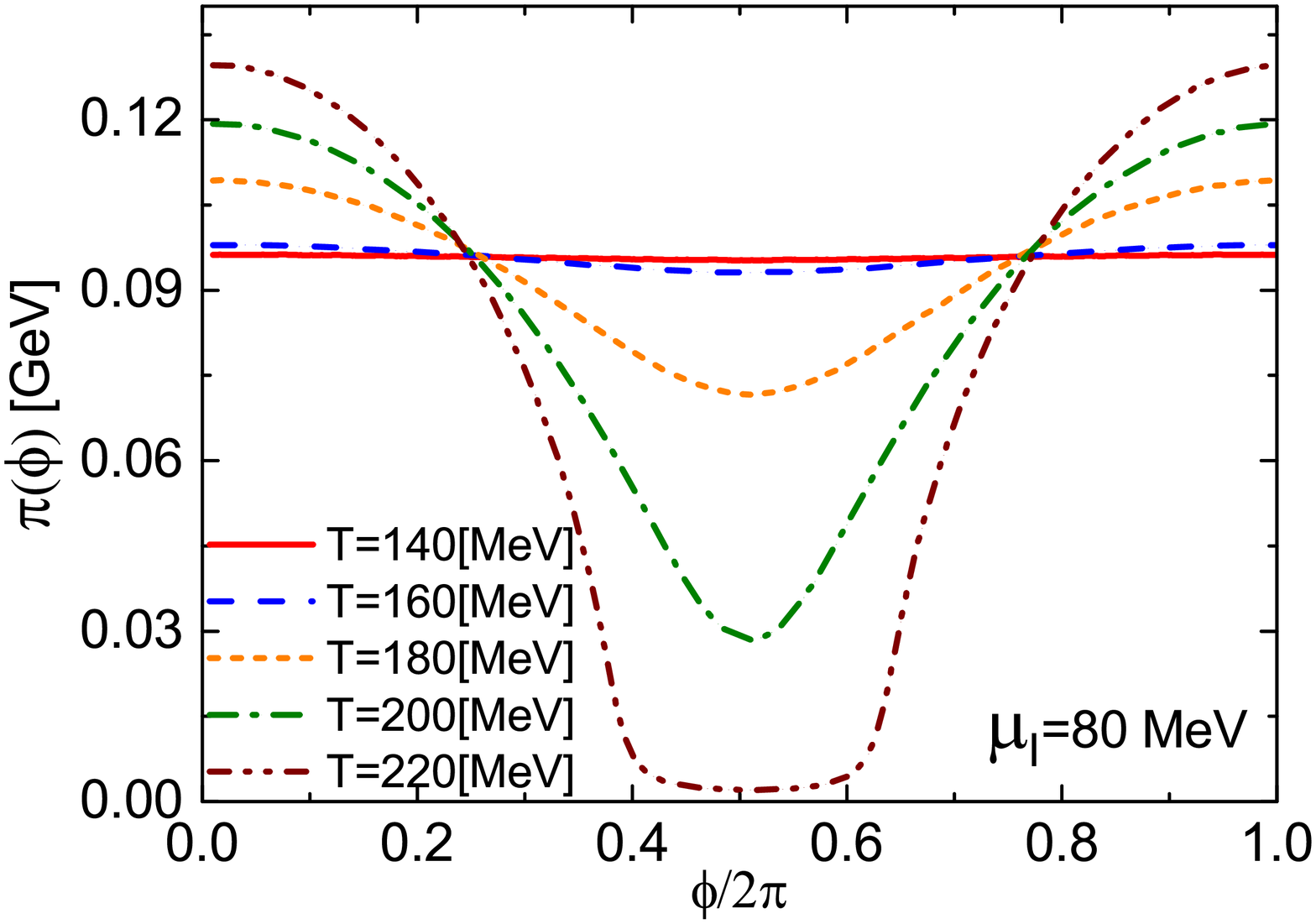}
\centerline{(b) }
\end{minipage}
\hspace{-.05\textwidth}
\begin{minipage}[t]{.45\textwidth}
\includegraphics*[width=\textwidth]{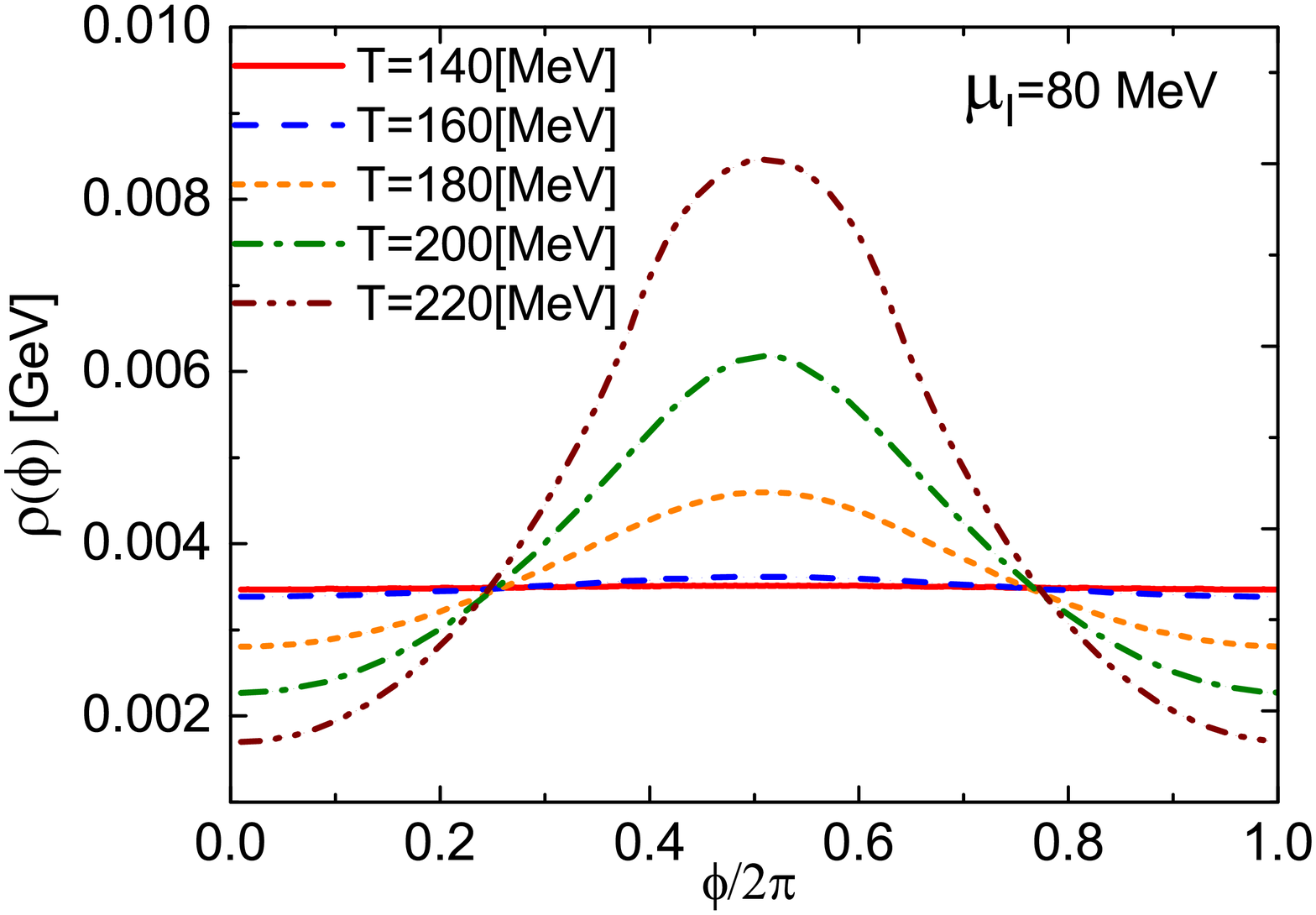}
\centerline{(c) }
\end{minipage}
\caption{ The twisted angle dependences of $\sigma$, $\pi$ and $\rho$ at $\mu_I=80 \MeV$
for different temperatures in PL$\sigma$M. The Dirac-sea contribution is included. }
\label{fig:dualc2}
\end{figure}

\begin{figure}[t]
\hspace{-.0\textwidth}
\begin{minipage}[t]{.45\textwidth}
\includegraphics*[width=\textwidth]{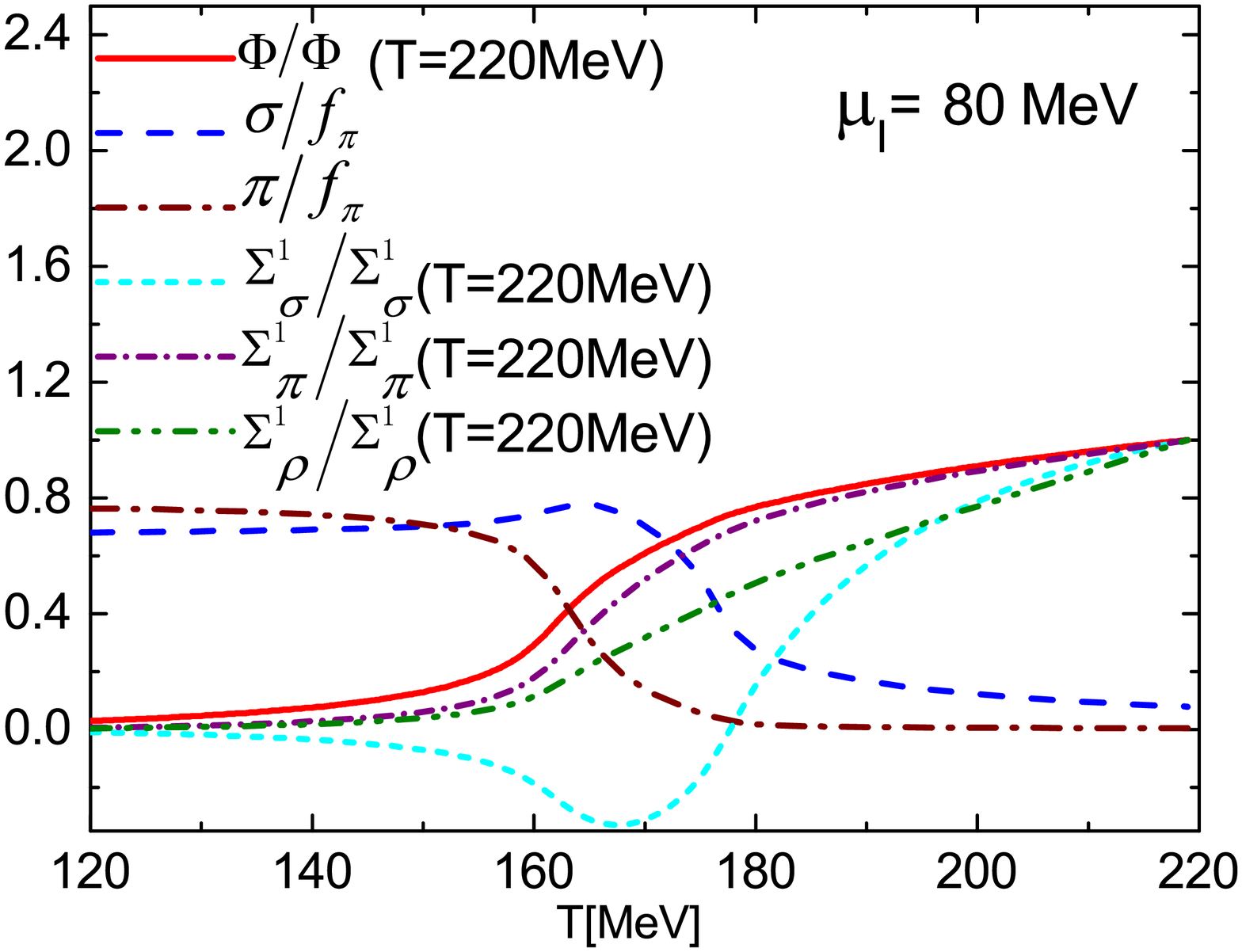}
\end{minipage}
\hspace{-.05\textwidth}
\begin{minipage}[t]{.45\textwidth}
\includegraphics*[width=\textwidth]{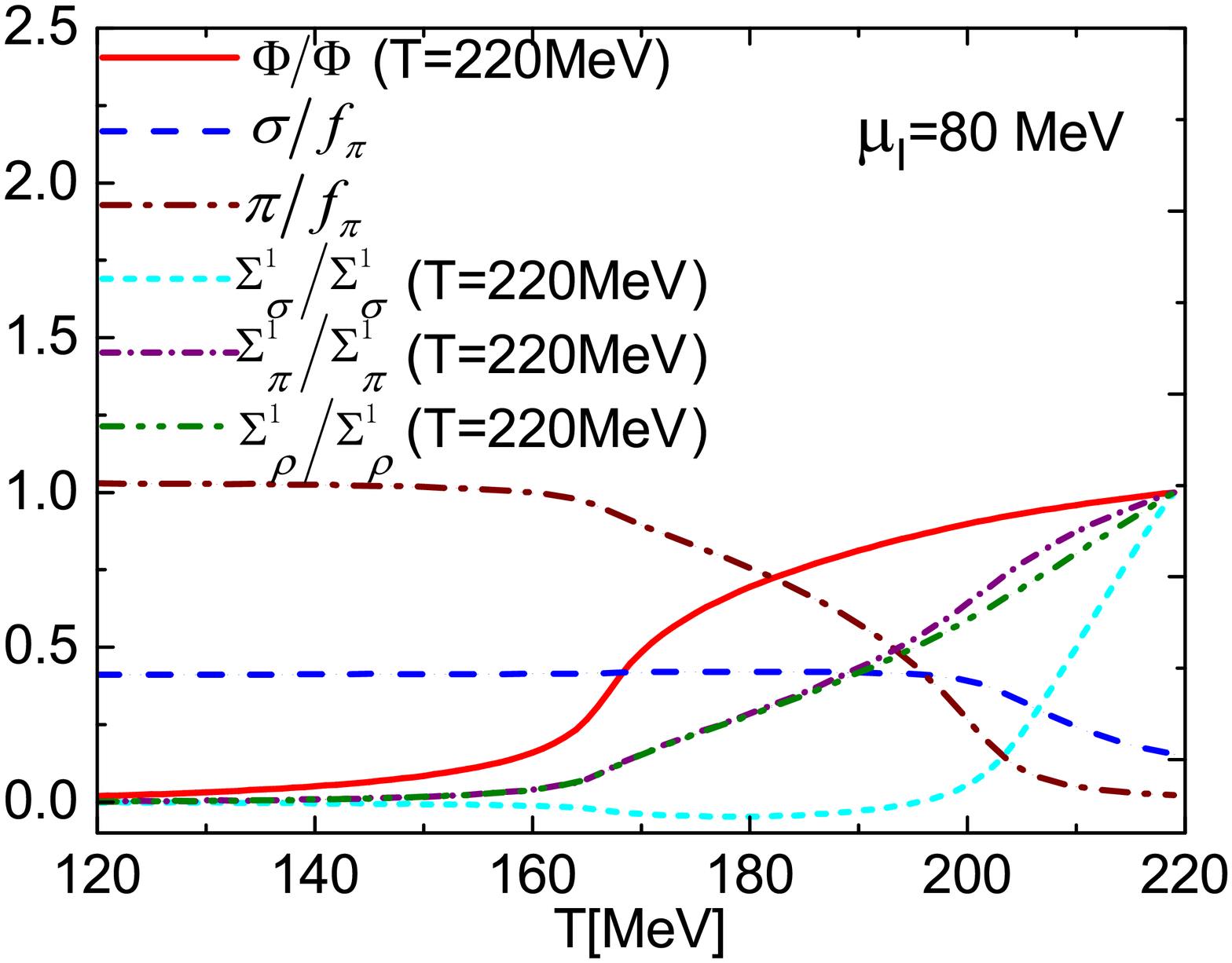}
\end{minipage}
\caption{ The temperature dependences of the normalized conventional Polyakov-loop,
sigma condensate, pion condensates, isospin density and their corresponding dual parters
at $\mu_I=80 \MeV$ in PL$\sigma$M. The upper (lower) panel is obtained by ignoring (considering) the Dirac-sea contribution. }
\label{fig:plcondensate}
\end{figure}

\begin{figure}[t]
\hspace{-.0\textwidth}
\begin{minipage}[t]{.45\textwidth}
\includegraphics*[width=\textwidth]{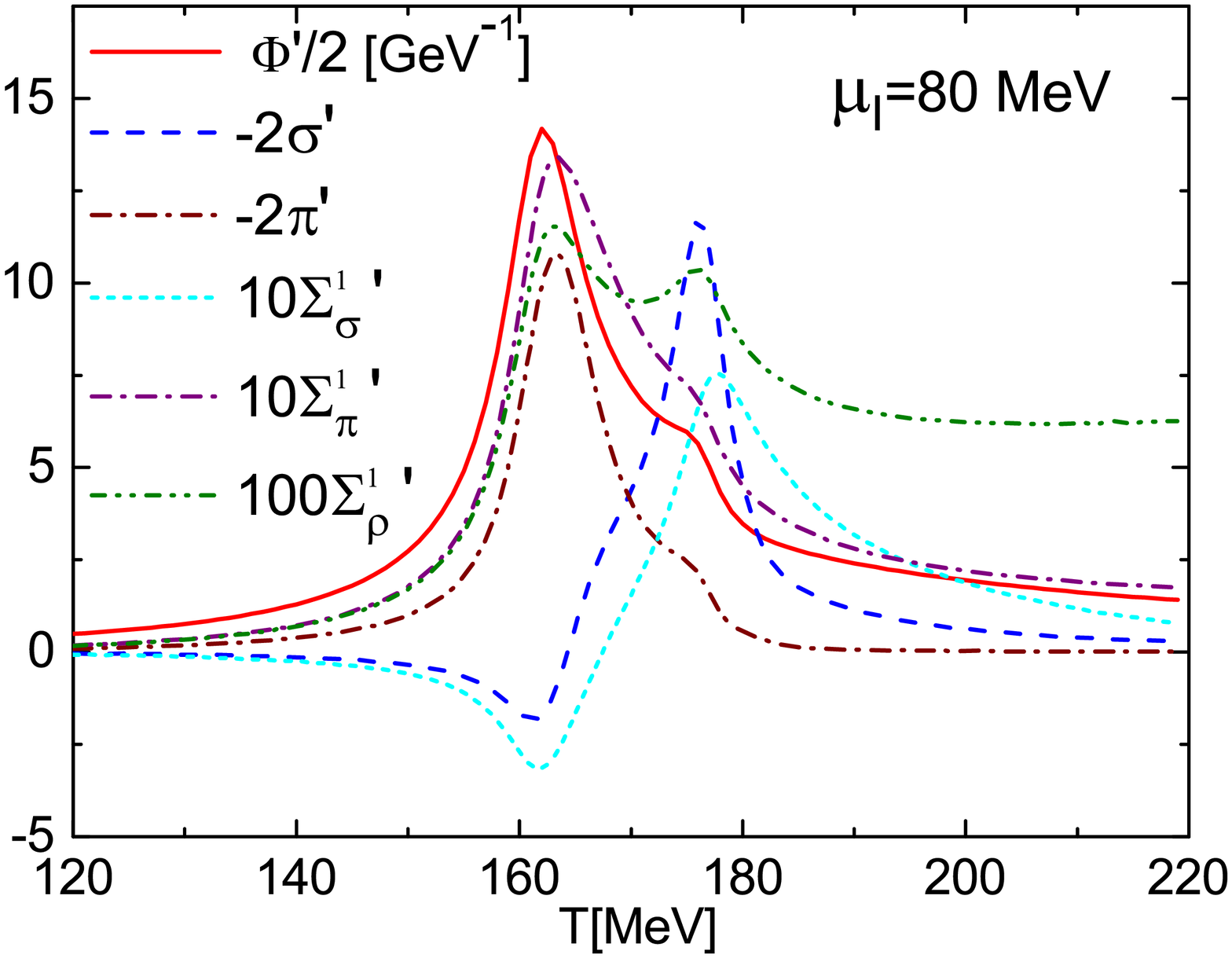}
\end{minipage}
\hspace{-.05\textwidth}
\begin{minipage}[t]{.45\textwidth}
\includegraphics*[width=\textwidth]{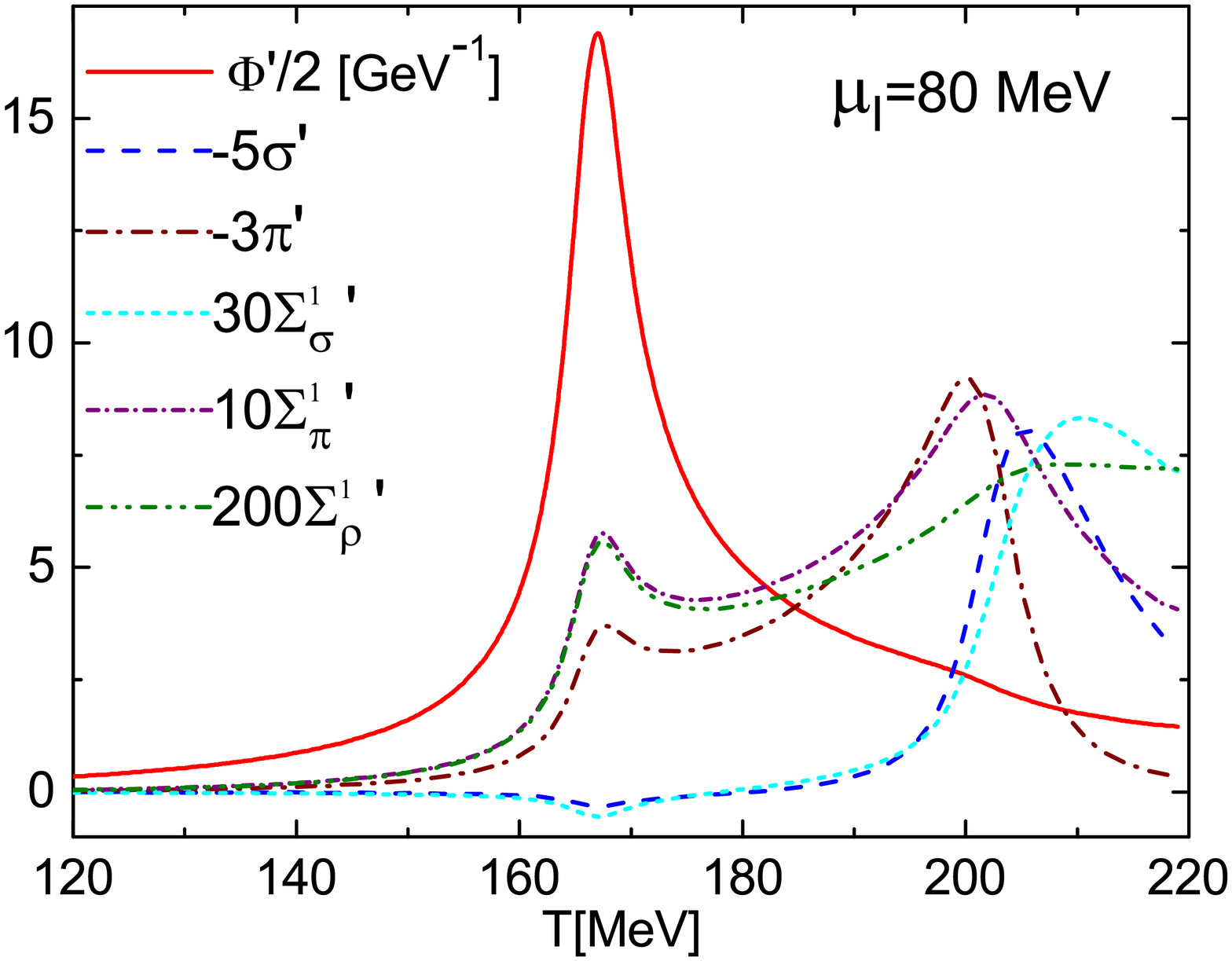}
\end{minipage}
\caption{ T-derivatives of the conventional Polyakov-loop, sigma and
pion condensates, dual sigma and pion condensates, dual isospin density
as functions of $T$ at $\mu_I=80 \MeV$ in PL$\sigma$M . The upper (lower) panel is obtained by ignoring (considering)
the Dirac-sea contribution. }
\label{fig:sus}
\end{figure}

\section{ Numerical results and discussions }

\subsection{ Dual sigma condensate for zero $\mu_I$ in PL$\sigma$M }

The thermal properties of D$\sigma$C and its relation with $\sigma$ and PL are first investigated at zero $\mu_I$ in
PL$\sigma$M. The results obtained by ignoring (including) the Dirac-sea contribution are shown in the upper (lower) panel
of each figure in this subsection.

\subsubsection{ $\phi$-dependence of the sigma condensate }

Figure.\ref{fig:sigmazeromi} shows the generalized sigma condensate as a function of the twisted angle $\phi$ at different fixed
temperatures. We see that at low temperatures, $\sigma$ is insensitive to $\phi$ and the line of $\sigma(\phi)$ is almost flat in
both panels. With increasing $T$, $\sigma$ decreases in the fermion-like region ($\phi\sim\pi$) but increases in the boson-like
region ($\phi\sim{0}$ or $2\pi$). These features are qualitatively consistent with that of the generalised quark condensate calculated
in LQCD \cite{Bilgici:2008qy}, truncated Dyson-Schwinger method \cite{Fischer:2009gk}, and  PNJL \cite{Kashiwa:2009ki}.

\subsubsection{ Thermal property of the dual sigma condensate }

The normalised D$\sigma$C, PL and $\sigma$ as functions of $T$ are shown in Fig.\ref{fig:dualzeromi}. Both panels display that D$\sigma$C
increases monotonically with $T$. We notice it is quite small at low temperature and raises rapidly in the chiral transition
region, no matter whether the Dirac-sea contribution is included or not. This means D$\sigma$C in PL$\sigma$M really behaves like DPL
obtained in LQCD and other methods at zero $\mu$.

The $T$-derivatives of quantities shown in Fig.\ref{fig:dualzeromi} are displayed in Fig.~\ref{fig:suszeromi}. In this paper, the
slope maximum is used to identify the critical temperature. Fig.~\ref{fig:suszeromi} indicates that the $T$-derivative of PL has
only one peak, which indicating the deconfinement temperature $T_c^P$. Differently, each of the slopes of $\sigma$ and D$\sigma$C
has double peaks: the former locates at $T_c^P$ and $T_c^{\chi}$, respectively, and the latter $T_c^P$ and $T_c^{d\sigma}$. Both panels
in this figure show that the slope maximum of D$\sigma$C is at $T_c^{d\sigma}$ rather than $T_c^{P}$ and $T_c^{d\sigma}\approx{T_c^{\chi}}$.
The coincidence of $T_c^{d\sigma}$ and $T_c^{\chi}$ implies D$\sigma$C is more sensitive to the drop of $\sigma$ rather than the
increase of PL.

We confirm that D$\sigma$C in L$\sigma$M shows the similar $T$-dependence, even no center symmetry is considered.
In this case, the $T$-derivative of D$\sigma$C peaks exactly at $T_c^{\chi}$ (in the chiral limit). This implies that the rapid rise
of D$\sigma$C is also totally driven by the chiral transition. Since D$\sigma$C corresponds to DPL, we conclude that the main result
in \cite{Benic:2013zaa,Marquez:2015bca} that DPL in NJL type models without center symmetry only reflects the chiral transition
is also supported by L$\sigma$M with quarks fields. Figs.~\ref{fig:dualzeromi}-\ref{fig:suszeromi} demonstrate that such a conclusion
is still valid in PL$\sigma$M.

\subsection{ Dual meson condensates for nonzero $\mu_I$ in PL$\sigma$M }

We then extend the study to $\mu_I>m_\pi/2$ to check whether the PNJL result in \cite{Zhang:2015baa} holds in
PL$\sigma$M too. Besides D$\sigma$C, the thermal properties of D$\pi$C and D$\rho$C are also investigated.

\subsubsection{ $\phi$-dependences of the meson condensates }

The generalized meson condensates $\sigma$, $\pi$ and $\rho$ as functions of $\phi$ for $\mu_I=80\MeV$
at different temperatures are shown in Fig.~\ref{fig:dualc1}, where the quark Dirac-sea contribution is
ignored.

Figure.~\ref{fig:dualc1}.a displays that in the fermion-like region, $\sigma(\phi)$ is a concave line for
$T>T_c^{I_3}$ (the melting temperature of pion superfluidity), but it becomes a convex one for $T<T_c^{I_3}$. This
is distinct with Fig.~\ref{fig:sigmazeromi}.a, where only concave curves emerge. The difference can be traced back
to the fact that $\sigma(\phi\sim\pi)$ first increases and then decreases with $T$ near $T_c^{I_3}$
\footnote{This is also observed in \cite{Zhang:2006gu} and other chiral model studies \cite{He:2005nk,Zhang:2006dn}. The
reason for such an anomaly is that comparing to the decline of $\sqrt{\sigma^2+\pi^2}$, $\pi$ drops more
significantly with $T$ near $T_c^{I_3}$. }. Moreover, this panel shows $\sigma$ decreases with $T$ near $\phi=0$,
which is also different from Fig.~\ref{fig:sigmazeromi}.a.

Figure.~\ref{fig:dualc1}.b shows that the $\phi$-dependence of $\pi$ is quite analogous to that of $\sigma$ displayed
in Fig.~\ref{fig:sigmazeromi}.a. The similarity can be understood in this way: For $\mu_I>m_\pi/2$, the sigma
condensate partially turns into the pion condensate, and thus the later inherits some properties of the former. Such a
transformation also leads to an obvious modification of $\sigma(\phi)$, as demonstrated in Fig.~\ref{fig:dualc1}.a.

The $\phi$-dependence of $\rho$ is shown in Fig.~\ref{fig:dualc1}.c. This quantity is also insensitive to $\phi$ at
low temperatures. With increasing $T$, it decreases near $\phi\sim\pi$ but increases around $\phi\sim{0}$. Thus only
convex lines appear in Fig.~\ref{fig:dualc1}.c.

Figure.~\ref{fig:dualc2} shows the same quantities obtained by including the Dirac-sea contribution. We see that the
$\phi$-dependence of $\sigma$ in the fermion-like region is analogous to that in Fig.~\ref{fig:dualc1}.a. In the
boson-like region, $\sigma$ increases with $T$ for a fixed $\phi$, which is different from Fig.~\ref{fig:dualc1}.a.
The $\phi$-dependences of $\pi$ and $\rho$ are all similar to that in Fig.~\ref{fig:dualc1}.

Note that as functions of $\phi$, $\pi$ and $\sigma$ for $\mu_I>m_\pi/2$ in PL$\sigma$M with the vacuum contribution
are qualitatively consistent with the corresponding PNJL results in \cite{Zhang:2015baa}.

\subsubsection{ Thermal behaviors of dual meson condensates }

Three dual meson condensates and other three (pseudo-) order parameters as functions of $T$ for $\mu_I=80\,\text{MeV}$
are shown in Fig.~\ref{fig:plcondensate}. The corresponding $T$-derivatives are displayed in Fig.~\ref{fig:sus}.
The numerical results obtained by ignoring (considering) the Dirac-sea contribution are still exhibited in the
upper (lower) panel of a figure.

Figure.~\ref{fig:plcondensate}.a indicates that $\pi$ ($\Phi$) decreases (increases) monotonically with $T$, but
$\sigma$ first increases with $T$ up to $T\sim170\,\text{MeV}$ and then decreases. Similar T-dependences are
observed in Fig.~\ref{fig:plcondensate}.b ($\sigma$ raises very slowly up to $T\sim200\,\text{MeV}$ and then declines).
As mentioned, the unnatural thermal behavior of  $\sigma$ is due to the fast dropping of pion condensate.
Fig.~\ref{fig:plcondensate} shows that D$\pi$C and D$\rho$C really behave like PL. In contrast, D$\sigma$C first
decreases with $T$ and then increases, which is quite different from PL, D$\pi$C, and D$\rho$C.

The abnormal $T$-dependence of D$\sigma$C can be attributed to the non-concave lines of $\sigma(\phi)$ displayed
in Figs.\ref{fig:dualc1}-\ref{fig:dualc2}, or the unusual T-dependence of $\sigma$ mentioned above. Actually,
Fig.~\ref{fig:plcondensate} clearly shows that when $\sigma$ increases with $T$, D$\sigma$C decreases, and vice versa.
This is further evidence that D$\sigma$C is quite sensitive to $\sigma$ but not PL, since the later always increases
with $T$. Such an anomaly is in agreement with the PNJL result in \cite{Zhang:2015baa}, where DPL exhibits the similar
thermal behavior.

Accordingly, the maximum of the D$\sigma$C slope still locates around $T_c^\chi$, as shown in Fig.~\ref{fig:sus}. The upper
panel of this figure displays each of the slopes of $\pi$, D$\pi$C, and PL has only one peak and the corresponding critical
temperatures $T_c^{I_3}$, $T_c^{d\pi}$, and $T_c^P$ almost have the same value. Note that here $T_c^{d\pi} \approx T_c^P$ is
just a coincidence. Actually, the lower panel shows that when taking into account the Dirac-sea contribution, $T_c^{d\pi}$ determined
by the maximum of the D$\pi$C slope is considerably larger than $T_c^P$, but $T_c^{d\pi}$ still equals to $T_c^{I_3}$. So we
conclude that $T_c^{d\pi}$ just denotes the melting temperature of pion condensate, even D$\pi$C behaves like PL. This
is also in agreement with the PNJL result \cite{Zhang:2015baa}.

Figure.~\ref{fig:sus} displays that the $T$-derivative of D$\rho$C peaks near $T_c^P$ and $T_c^\chi$, respectively. The upper
panel indicates  $T_c^\rho \approx T_c^P$,  but the lower one exhibits $T_c^\rho \approx T_c^\chi$. Here $T_c^\rho$ denotes the
location of the highest peak. This implies D$\rho$C is more sensitive to $\sigma$ (PL) with (without) the Dirac-sea
contribution.
We have checked that the result $T_c^\rho \approx T_c^\chi$ is supported by PNJL.
We thus argue that $T_c^\rho \approx T_c^P$ should be an artifact of PL$\sigma$M without the vacuum contribution.

\subsection{ Discussions }

Our calculations suggest that the slope of each dual meson condensate exhibits double peaks in PL$\sigma$M
and the lower one is determined by PL. This is similar to the $T$-derivative of the corresponding meson condensate.
In this sense, the dual meson condensates obtained in the two-flavor L$\sigma$M of QCD are not qualified
order parameters for deconfinement, even the center symmetry is considered. This conclusion is consistent with
NJL studies \cite{Benic:2013zaa,Marquez:2015bca,Zhang:2015baa}.

The similar results in PL$\sigma$M and PNJL may be indicative for QCD. First, the center symmetry is severely violated.
So it is very likely that some dual observables, such as DPL or D$\sigma$C, are insensitive to deconfinement or PL, unless
the dynamical quarks are heavy enough. Second, formally, the definition of DPL (D$\sigma$C) is naturally related to the
quark (sigma) condensate. Thus it is not strange that DPL (D$\sigma$C) is more sensitive to the chiral transition.
Such a viewpoint is supported by the recent study of Dirac-mode expansion at imaginal chemical potential \cite{Doi:2017dyc}:
it shows that even VEVs of some quark bilinears can be expressed as PL and its conjugate for large quark mass, the quark
number density (also the quark condensate) is still strongly dependent on low-lying Dirac-modes for small quark mass
\footnote{ It is reported in \cite{Doi:2017dyc} that the sign of the quark number density is insensitive to low-lying
Dirac-modes, which supports the quark number holonomy \cite{Kashiwa:2016vrl} as the deconfinement indicator.}.
Thus it might be misleading to conclude the coincidence of chiral restoration and deconfinement through studying
DPL (D$\sigma$C).

Of course, PL$\sigma$M and PNJL are just simple models which may only partially reflect the possible relation between the
(dynamically) center symmetry breaking and a dual observable existing in QCD. So DPL or D$\sigma$C mainly indicates the
chiral transition in \cite{Benic:2013zaa,Marquez:2015bca,Zhang:2015baa} and our calculation may not really happen in QCD.
In addition, the investigations in PL$\sigma$M and PNJL do not exclude the possibility that some dual observables may be
sensitive to deconfinement but insensitive to chiral transition.

Generaly, deconfinement is associated with the liberation of degrees of freedom, manifested by the rapid rise in bulk
thermodynamical quantities, such as the pressure, energy density, etc. Among them, the appropriate combinations of
fluctuations and correlations of different conserved quantum numbers, for example, $\chi^B_2-\chi^B_4$ and
$\chi^{BS}_{31}-\chi^{BS}_{11}$, directly probe the liberation of quark degrees of freedom \cite{Bazavov:2013dta,Bellwied:2013cta}.
It is interesting to study whether dual observables constructed from these bulk thermodynamical quantities are sensitive
to deconfinement. There is a discussion on the dual pressure as the order parameter in \cite{Kashiwa:2016btd}. Further
investigation on this topic is needed which is beyond the scope of this paper.

Note that recent lattice calculations \cite{Bazavov:2013yv,Bazavov:2016uvm} show that in the temperature region where
the quark condensate drops rapidly, the renormalized PL is still quite small ($\sim$0.1 near $T_c^\chi$) and changes
quite mildly. This implies there is no obvious connection between the chiral and deconfinement transitions described in
terms of these quantities. In contrast, PL calculated in effective models is relatively large near $T_c^\chi$, which reaching
unity quickly. This discrepancy has been discussed recently by Pisarski and Skokov in the chiral matrix model \cite{Pisarski:2016ixt}
{\footnote{ $T_c^P$ extracted from PL is almost the same as $T_c^\chi$ in this model.}} and the reason is still unclear.
On the other hand, the entropy of static quark calculated in lattice simulation \cite{Bazavov:2016uvm} suggests that the
deconfinement and chiral transitions happen in the similar temperatures. Thus whether the chiral transition and deconfinement
have a close relation or not is still a subtle problem and the sensitive probe of deconfinement needs to be further
investigated.

\section{ Conclusion }

Dual meson condensates as possible order parameters for center symmetry are tested in PL$\sigma$M. We mainly focus on
the thermal property of the dual sigma condensate. The dual pion and vector meson condensates at $\mu_I>m_{\pi}/2$ are
also investigated. To our knowledge, this is the first paper for employing PL$\sigma$M at imaginal chemical potential.

At zero density, we find that D$\sigma$C really behaves like the thin or dressed PL. Its rapid rise with $T$ near $T_c^{\chi}$
is driven by the drop of $\sigma$ rather than the increase of PL. So the critical temperature determined by D$\sigma$C
just indicates the chiral transition rather than deconfinement. It is confirmed that L$\sigma$M without center symmetry
gives the similar result.

For $\mu_I>m_{\pi}/2$, D$\sigma$C shows abnormal thermal behavior: it first decreases with $T$ and then increases, which
is distinct with PL. We reveal that D$\sigma$C increases with $T$ when $\sigma$ decreases, and vice versa. The anomaly is
a further evidence that D$\sigma$C is quite sensitive to the chiral dynamics but insensitive to the center symmetry. In
contrast, D$\pi$C and D$\rho$C still exhibit the similar $T$-dependence as PL. We verify that
the maximum slope of D$\pi$C does not indicate deconfinement, but the restoration of $U(1)_{I_3}$ symmetry. Analogously, when
taking into account the Dirac-sea contribution, the rapid rise of D$\rho$C is driven not by deconfinement but by the chiral
restoration.

We thus conclude that the dual meson condensates are not appropriate order parameters for deconfinement in PL$\sigma$M
(also in L$\sigma$M). Our results are qualitatively consistent with the calculations of NJL at zero density
\cite{Benic:2013zaa,Marquez:2015bca} and PNJL at $\mu_I>m_{\pi}/2$ \cite{Zhang:2015baa}. We argue that the reason can be
attributed to either the fact that the center symmetry is seriously broken by light quarks and thus not all the dual
observables are qualified order parameters for deconfinement or the limitation of simple models in which some intrinsic
connection between the center symmetry and a dual observable is ignored.

\vspace{5pt}
\noindent{\textbf{\large{Acknowledgements}}}\vspace{5pt}\\
Z.Z. was supported by the NSFC ( No.11275069 ).

\end{document}